\documentstyle[preprint,pre,aps,eqsecnum]{revtex}

\begin{document}
\draft

\title{The lattice gas model with isospin-dependent interactions}

\author{J. Pan and S. Das Gupta}

\address{
Physics Department, McGill University,
3600 University St., Montr{\'e}al, Qu{\'e}bec \\ Canada H3A 2T8}


\maketitle

\begin{abstract}
In this paper we continue the investigation of the lattice gas model.
The main improvement is that we use two strengths for bonds: one between 
like particles and another between unlike particle to implement the 
isospin-dependence of nuclear force.  The main effect is the elimination
of unphysical clusters, like the dineutron or diproton. It is therefore a
better description of nuclear system. Equation of state in mean field
theory is obtained for nuclear matter as well as for $N\ne Z$ 
systems.  Through numerical and analytical calculation we show that the 
new model maintains all the important features of the older model. 
We study the effect of the Coulomb interaction on multifragmentation of a 
compound system of A=86, Z=40 and also for A=197, Z=79.  For the first
case the Coulomb interaction has small effect.  For the latter case the effect
is much more pronounced but typical signatures of the lattice gas model such as
a minimum (maximum) in the value of $\tau$ ($S_2$) are still obtained but
at a much lower temperature. 
\end{abstract}


\newpage
\section{Introduction}

Over the past a few years we have been developing a lattice gas model 
for the study of nuclear 
multifragmentation[1-5].  In this model $n$ nucleons are placed in $N$ cubes
and they interact with nearest neighbour interactions.  In most of this work
the interaction between neutron-neutron, proton-proton and neutron-proton
was  taken to be identical although we have sometimes used[5] a more
complicated model in which interactions between like particles are different
from those between unlike particles.  The purpose of this paper is to more
fully expose this improved model and to examine its relationship with our
earlier and simpler model.  Among other things we will also show that the
important conclusions reached with our simpler model go through in our
improved version.

The lattice gas model has the attractive feature that it can provide an
equation of state but it can also provide the cluster distribution.  Most
of the observables are calculated using Monte-Carlo simulations.
We used Metropolis algorithm in our simulations.  The $n$
particles are distributed in $N$ boxes according lattice gas Hamiltonian 
and their momenta are generated from Maxwell-Boltzmann distribution 
at a prescribed temperature.
Calculation of clusters is  straightforward.  Two nucleons in neighbouring
cells are part of the same cluster if the relative kinetic energy is less
than the strength of the attractive bond: $p_r^2/2\mu+\epsilon<0$.  Here
$p_r$ is the relative momentum, $\mu$ the reduced mass and $\epsilon$ is
negative(attractive interaction).  This prescription is sufficient to calculate
cluster distribution.  The motivation for introducing two kinds of bond is
now obvious.  If the neutron-neutron bond or the proton-proton bond is
attractive then each numerical simulation can generate
dineutrons and diprotons which in reality do not exist in nature as
composites.  One can get rid of these unphysical clusters by simply making
the neutron-neutron and proton-proton bonds zero or repulsive.

For completeness we mention some more features of the lattice gas model that
were studied in [1,2].
  One finds that at a certain temperature the distribution of composites
is a power law: $Y(Z)\propto  Z^{-\tau}$ where $Y(Z)$ is the number 
(averaged over many simulations) of composites with $Z$ protons.  As is a common
practice, we will extract a value of $\tau$ even when the distribution has
significantly deviated from a power law[6].  This value is extracted by using
\begin{eqnarray}
\frac{\sum_2^{10}ZY(Z)}{\sum_2^{10}Y(Z)}=\frac{\sum_2^{10}ZZ^{-\tau}}
{\sum_2^{10}Z^{-\tau}}
\end{eqnarray}
We also calculate the second moment $S_2=\sum'A^2Y(A)/n$ where in the sum
the largest cluster is excluded.  The usefulness of $S_2$ was emphasized by
Campi[7].  Since the lattice gas model has a Hamiltonian, its average energy
at any temperature can be calculated.  The excitation energy and specific
heat $C_v$ per particle can be calculated.  We will find this useful too.

\section{The equation of state}

We have $N$ boxes in the lattice in which we have to put $n_n$ neutrons 
and $n_p$ protons;
$n_p+n_n=n<N$; $n/N=V_0/V=\rho/\rho_0$ where $V_0$, $\rho_0$ are
normal nuclear volume and density respectively.  
In principle, we can have three kinds of bonds: $\epsilon_{nn},
\epsilon_{pp}$ and $\epsilon_{pn}$.  In the most general case where these
interactions can take any arbitrary values a rich assortment of phenomena is
predicted.  The grand canonical partition function  of this general lattice gas
model can be mapped on to a spin 1 Ising type model in the presence of a 
magnetic and quadrupole field.  These have been studied in detail in the
past[8].  For the nuclear case the values of the interactions are quite
restricted and the richness of phenomena disappears.  First of all we have to
set $\epsilon_{nn}$ and $\epsilon_{pp}$ to be either zero or repulsive so
that one avoids producing unphysical dineutron or diproton bound clusters.
Charge independence of nuclear forces suggests that we put $\epsilon_{nn}=
\epsilon_{pp}$.  From now on we will write $\epsilon_{pp}$ for both
$\epsilon_{pp}$ and $\epsilon_{nn}$.  
In our past work[5] and also in other modelings[9] of nuclear
collisions using classical mechanics a slightly repulsive $\epsilon_{pp}$ was
used.  To avoid proliferation of parameters we will set this bond to
zero in this work.  The binding energy of nuclear matter fixes the value
of $\epsilon_{pn}$ at -5.33 MeV.  

Throughout this work $\gamma$ stands for the number of nearest neighbours.
In 3 dimensions one has $\gamma=6$.  We use the Bragg-Williams mean
field theory using
the canonical ensemble.  There are $N$ boxes and $n_p$ protons and $n_n$
neutrons.  Let one of the boxes be occupied by
a proton.  Then, in the Bragg-Williams approximation, among its nearest 
neighbours, on the average $\gamma n_p/N$ will be occupied by protons and 
$\gamma n_n/N$ by neutrons.  The number of $n-p$ bonds will be 
$\gamma n_pn_n/N$, the number of $p-p$ bonds will be 
$(1/2)\gamma n_pn_p/N$ where the 
factor of 1/2 remedies the double counting for proton-proton bonds.  Similarly
starting with a box occupied by a neutron we come up with the same number of
neutron-proton bonds and the number of neutron-neutron bonds is determined to
be $(1/2)\gamma n_nn_n/N$.  Thus the interaction energy when there are
$n_p$ protons, $n_n$ neutrons placed in $N$ boxes is
$E=\gamma (\epsilon_{pn}n_pn_n+\epsilon_{nn}(n_n^2+n_p^2)/2)/N$
and the partition function is
\begin{eqnarray}
Z(N,n_p,n_n)=\frac{N!}{(N-n_p-n_n)!n_p!n_n!}\exp (-\beta E)
\end{eqnarray}
We now find pressure $P$ from the equation $P=kT[\partial \ln Z/\partial V]_T$
and $V=a^3N$ where $a^3=1/\rho_0$ is the volume of each box.  Using Stirling's
formula one arrives at
\begin{eqnarray}
P=\rho_0 kT\ln\frac{N}{N-n}+\rho_0\gamma\epsilon_{pn}(n_p/N)(n_n/N)+
\rho_0\gamma\epsilon_{nn}((n_n/N)^2+(n_p/N)^2)/2
\end{eqnarray}
Introduce an asymmetry parameter $\eta=(n_n-n_p)/(n_n+n_p)$ which takes value
1 for neutron matter, 0 for nuclear matter and -1 for proton matter.  We
can then write
\begin{eqnarray}
P=\rho_0kT\ln\frac{V}{V-V_0}+\frac{1}{2}\rho_0\gamma\frac{V_0^2}{V^2}[
\frac{\epsilon_{pn}+\epsilon_{nn}}{2}+\frac{1}{2}\eta ^2(\epsilon_{nn}-\epsilon_
{pn})]
\end{eqnarray}
Determine the critical point from $\partial P/\partial \rho=\partial
^2P/\partial \rho^2=0$. This gives the critical density $\rho_c=.5\rho_0$
and the critical temperature to be 
$-(\gamma/4)[(\epsilon_{pn}+\epsilon_{pp})/2
+(1/2)\eta ^2(\epsilon_{nn}-\epsilon_{pn})]$.  In this approximation with 
 $\epsilon_{pn}$ attractive and $\epsilon_{nn}=0$ the critical temperature for
 nuclear matter (which has $\eta$=0) would be highest at 
$-(\gamma/4)(\epsilon_{pn}/2)$ and would fall
off quadratically with $\eta$ to 0 at neutron or proton matter.

The Bragg-Williams approximation  is the simplest mean-field approximation.
An improved treatment using Bethe-Peierls approximation is worked out in the
appendix.  The mean field calculation shown in this section and the appendix is
merely to form a rough idea about the nature of phase transition.  In
practical calculations we need to obtain the yields of the composites at a
given temperature.  Mean field theories do not provide these and we need to
do event by event calculation which can be obtained through Monte-Carlo
samplings.

\section{Monte-Carlo results}

  As far as we know exact results with two kinds of bonds are not
available.  For one kind of bond one can often interpret essentially exact 
although numerical results from well-studied spin 1/2 Ising model for use
in the lattice gas model.  In the absence of such exact results our only
recourse is to compare numerical results obtained with two kinds of bonds,
i.e., $\epsilon_{pn}=-5.33$ MeV, $\epsilon_{nn}=0$ with those obtained with
one kind of bond that we have used before, i.e., $\epsilon_{pn}=\epsilon_{nn}=
-5.33$ MeV.  We use here $N=7^3$, which is a number appropriate for finite 
systems that we will investigate.  As mentioned in the introduction the 
calculation proceeds by first putting the required number of protons and
neutrons in the $N$ boxes using a standard Metropolis algorithm.  Nucleons
are then assigned momenta from Monte-Carlo sampling of a Boltzmann distribution
at the given temperature.  The energy of the event can now be calculated. 
Clusters are then determined as explained in the introduction.  The results
shown in Figs. 1 and 2 are obtained by averaging over 1000 events for selected
temperatures.  For two assumed freeze-out densities
we calculate the specific heat, the second moment $S_2$ and the deduced values
of $\tau$.  The unit chosen in the graph for temperature is 
$T_c=1.1275|\epsilon_{pn}|$ which is the $T_c$ for an infinitely large lattice
with one kind of bond. We find that the peaking of $C_v, S_2$ and the minimum
of $\tau$ happen at a slightly lower temperature (about ten percent) with two 
kinds of bond as compared to when the same bond is used for all the particles.
For example for the minimum of $\tau$ to appear at the same temperature the
value of $\epsilon_{pp}=\epsilon_{pn}$ has to be set at about 10\% lower value
than the value of $\epsilon_{pn}$ when $\epsilon_{pp}$ is set to zero. 
Qualitatively and even semi-quantitatively the results in the two models look
similar when this renormalisation of the strength  is done.  However,
the peaking of $C_v$ is more pronounced in the two bonds model.

All models which employ freeze-out densities assume that the freeze-out density
is less than .5$\rho_0$.  If the freeze-out density is less than .5$\rho_0$
then in the lattice gas model a peak in the $C_v$ will signify the crossing 
of the co-existence curve and a first order phase transition.  The value
of specific heat can be deduced from the caloric curve [10] but locating the 
peak is very difficult in experiment.  In a recent paper we have suggested[11] 
that
since the peaking of $C_v$ is accompanied by a minimum in $\tau$ and a maximum
in $S_2$, the appearance of the last two could be taken as a signal of the
phase transition.  The appearance of the maxima in $S_2$ and of the minimum
in $\tau$ in close vicinity of the maximum in $C_v$ happens in both the versions
of the lattice gas model. 

\section{A study on the effects of the Coulomb force}

Here we follow the methods employed in ref [3].  At that time we studied the
influence of the Coulomb force on fragmentation of a system which had 85
nucleons and found the effects to be small.  For a much larger system
(Au on Au: central collision so that the compound system has $A\approx 394$)
we found the effect to be very large.  One of the rather unavoidable features
of the lattice gas model is the appearance of a minimum in the extracted value
of $\tau$ as a function of temperature.  This feature disappeared for the very
large system of $A=394$ because of the Coulomb interaction.

For completeness a short description of a similar calculation but done for
two kinds of bonds will be given here.  In addition to lattice gas calculations,
we do molecular dynamics calculations whose purpose is two-fold.  One is to
check if the predictions of a lattice gas model can resemble those of a
molecular dynamics calculation provided the initial conditions are the same and
the forces are chosen to be such that
they resemble implied forces of the lattice gas model.  For this we place the
$n_p$ protons and $n_n$ neutrons in the N boxes using as usual Metropolis
algorithm.  Next we assign the momenta from Monte-Carlo sampling of a
Maxwell-Boltzmann distribution.  Once this is done the lattice gas model
immediately gives the cluster distribution using the rule that two nucleons are
part of the same cluster if $p_r^2/2\mu+\epsilon<0$.  To calculate clusters
using molecular dynamics
we propagate the particles from this initial configuration for a long time 
under the influence of the chosen force (we will give the force parameters
shortly).  At asymptotic times the clusters are easily recognised (a detailed
discussion of cluster recognition which requires shorter computer times
can be found in ref 12).  The cluster
distribution in the two models can now be compared.  Fig. 3 shows the two
prescriptions give nearly the same answer.  

We now come to the second and more important purpose of the molecular dynamics
calculation.  We now add the Coulomb interaction to the nuclear part.  The
initialisation of putting the nucleons in $N$ boxes is done again but now with
the inclusion of Coulomb forces.  We then do a molecular dynamics propagation
including the Coulomb force.  The clusters can again be calculated and compared
with the cases where the Coulomb force was ignored.

We now give the force parameters for molecular dynamics propagation.  The
neutron-proton potential was taken to be
$v_{pn}(r)=A[B(r_0/r)^p-(r_0/r)^q]\exp ([1/(r/r_0-a)])$ for $r/r_0<a$
and $v_{pn}(r)=0$ for $r/r_0>a$.  Here $r_0=1.842$ fm is the distance 
between the
centers of two adjacent cubes.  We have chosen $p=2, q=1, a=1.3, B=.924$ and
$A=1966$ MeV.  With these parameters the potential is minimum at $r_0$ with
the value -5.33 MeV, is zero when the nucleons are more than $1.3r_0$ apart
and becomes strongly repulsive when $r$ is significantly less than $r_0$.
We now turn to the nuclear part of like particle interactions.  Although we
take $\epsilon_{pp}=0$ in lattice gas calculations the fact that we do not put 
two like particles in the same cube would suggest that there is short range
repulsion between them.  We have taken the nuclear force between two like
particles to be the same expression as above plus 5.33 MeV upto $r=1.842$
and zero afterwards: $v_{pp}(r)=v_{pn}(r)-v_{pn}(r_0)$ for $r<r_0$ and
0 afterwards.  This means there is a repulsive core which goes to zero at 
$r_0$ and is zero afterwards.

The results shown in Figs. 3 and 4 can be summarised as follows.  Fig 3. first
of all shows that if there is no Coulomb interaction then lattice gas
model results are quite close to that of molecular dynamics simulation provided
in the latter one starts from  the same initial condition and uses a force
suitably chosen.  Fig 3 also shows that in the case of $A=85,Z=40$ the Coulomb 
force does not have a large effect.  The minimum in $\tau$ and the maximum
in $S_2$ are shifted to slightly lower temperature.  The effect for $A=197,Z=79$
is much bigger.  The minimum in $\tau$ and the maximum in $S_2$ are shifted
from 4.8 MeV (lattice gas without Coulomb) to about 2.4 MeV.  Our previous
calculation showed that there is no minimum in $\tau$ for $A=394, Z=158$.  So 
somewhere between these two limits the minimum will vanish.

\section{Discussion}

The two bond model is a natural progression of the simpler lattice gas model.
In this paper we have done calculations with the two bond model.
Although in detail the two models differ the major characteristics of the
well studied simple model remain unchanged.  The lattice gas model remains a 
quick tool to calculate experimental data.

When the Coulomb force is very strong the lattice gas model can not be
relied upon.  Figs. 3 and 4 give some indication of the reliability of the
model in the presence of a Coulomb force.  Calculations above indicate that a 
viable (although much more time consuming) prescription might be: obtain
the initial conditions as in a lattice gas model.  Put $n$ nucleons in $N$ boxes
by Metropolis sampling where one includes in addition to lattice gas Hamiltonian
the Coulomb force.  Obtain momenta of each nucleon from a Monte-Carlo 
sampling of a Maxwell-Boltzmann distribution.  Then propagate by molecular
dynamics to obtain cluster distributions.  Techniques developed in ref. 12
might be useful so that one does not need to run molecular dynamics till
asymptotic times.  One attractive feature of this hybrid model is that the
Coulomb force is operative even during the formation of clusters as opposed
to other models where the Coulomb force only adds repulsion between the
composites already formed.

\section{acknowledgements}
This work is supported in part by the Natural Sciences and Engineering Council
of Canada and by le fonds pour la formation de Chercheurs et l'aide \`a la
Recherche du Qu\'ebec.  We like to thank W. F. J. M\"uller for informing us
about reference 8.   

\pagebreak
\appendix
\section{}

We follow the method of reference 1.  We break up the lattices into
$N/(\gamma +1)$ blocks, each of which contains 1 central box and $\gamma$
nearest neighbours to it.  We refer to Fig A.1 where for simplicity a
two-dimensional lattice is shown.  The interactions within each block are
taken into account exactly while the interactions between different blocks
are treated in an approximate fashion.  The grand partition function can be
written as the product of the grand partition functions of the $N/(\gamma+1)$
blocks:
\begin{eqnarray}
Z_{gr}=z_{gr}(block 1)z_{gr}(block 2)...........z_{gr}(block\frac{N}{\gamma+1})
\end{eqnarray}
We want to write down the grand partition function of the block denoted by
1,2,3,$\gamma$ and 5.  In the general case there will be two absolute
fugacities $\lambda_p$ and $\lambda_n$, the first referring to protons and
the second to neutrons.  The grand partition function for a block can be
written as
\begin{eqnarray}
z_{gr}=A+B+C
\end{eqnarray}
where
\begin{eqnarray}
A=(1+e^{\lambda_p-\beta\bar{\epsilon}_p}+e^{\lambda_n-\beta\bar{\epsilon}_n})^
{\gamma}
\end{eqnarray}
\begin{eqnarray}
B=e^{\lambda_p}(1+e^{\lambda_p-\beta\bar{\epsilon}_p-\beta\epsilon_{pp}}+
e^{\lambda_n-\beta\bar{\epsilon}_n-\beta\epsilon_{pn}})^{\gamma}
\end{eqnarray}
\begin{eqnarray}
C=e^{\lambda_n}(1+e^{\lambda_p-\beta\bar{\epsilon}_p-\beta\epsilon_{pn}}+
e^{\lambda_n-\beta\bar{\epsilon}_n-\beta\epsilon_{pp}})^{\gamma}
\end{eqnarray}
Here $A$=contribution to the partition function when the central (innermost)
site is empty, $B$=contribution to the partition function when the central site
has a proton and $C$=contribution to the partition function when the central
site has a neutron.  Equation (4) takes the place of eq.(3.9) in ref [1].
The two constants $\bar{\epsilon}_p$ and $\bar{\epsilon}_n$ are the average
interaction energy with the adjacent block when a proton (neutron) occupies
a peripheral site.

The probability that the central site is occupied by a proton is $n_p/N$.  Thus
we have 
\begin{eqnarray}
\frac{n_p}{N}=\frac{B}{z_{gr}}
\end{eqnarray} 
But since no particular site is favoured over another one, the average 
occupation occupation of one of the peripheral sites must also be $n_p/N$.
This gives
\begin{eqnarray}
\frac{n_p}{N}=\frac{E+F+G}{z_{gr}}
\end{eqnarray}
where
\begin{eqnarray}
E=e^{\lambda_p-\beta\bar{\epsilon}_p}(1+e^{\lambda_p-\beta\bar{\epsilon}_p}
+e^{\lambda_n-\beta\bar{\epsilon}_n})^{\gamma-1}
\end{eqnarray}
\begin{eqnarray}
F=e^{\lambda_p}e^{\lambda_p-\beta\bar{\epsilon}_p-\beta\epsilon_{pp}}
(1+e^{\lambda_p-\beta\bar{\epsilon}_p-\beta\epsilon_{pp}}+
e^{\lambda_n-\beta\bar{\epsilon}_n-\beta\epsilon_{pn}})^{\gamma-1}
\end{eqnarray}
\begin{eqnarray}
G=e^{\lambda_n}e^{\lambda_p-\beta\bar{\epsilon}_p-\beta\epsilon_{pn}}
(1+e^{\lambda_p-\beta\bar{\epsilon}_p-\beta\epsilon_{pn}}+
e^{\lambda_n-\beta\bar{\epsilon}_n-\beta\epsilon_{pp}})^{\gamma-1}
\end{eqnarray}
Similarly two equations can be written for $n_n/N$.
\begin{eqnarray}
\frac{n_n}{N}=\frac{C}{z_{gr}}
\end{eqnarray}
\begin{eqnarray}
\frac{n_n}{N}=\frac{I+J+K}{z_{gr}}
\end{eqnarray}
where $I,J,K$ can be written down from expressions $E,F,G$ by interchanging
protons with neutrons.

Equations (8),(9),(13)and (14) determine the four constants $\lambda_p,\lambda_n
,\bar{\epsilon}_p,\bar{\epsilon}_n$.

For $n_p=n_n$, the calculations simplify.  Now we have $\lambda=\lambda_n=
\lambda_p$ and $\bar{\epsilon}=\bar{\epsilon}_n=\bar{\epsilon}_p$. Then
$B=E+F+G$ (eqs(8) and (9)) leads to
\begin{eqnarray}
e^{\lambda}(1+Qe^{\lambda-\beta\bar{\epsilon}})^{\gamma-1}=
e^{\lambda-\beta\bar{\epsilon}}(1+2e^{\lambda-\beta\bar{\epsilon}})^{\gamma-1}
+e^{2\lambda}(1+Qe^{\lambda-\beta\bar{\epsilon}})^{\gamma-1}e^{\beta\bar
{\epsilon}}Q
\end{eqnarray}
where we have defined $Q=e^{-\beta_{pp}}+e^{-\beta_{pn}}$.  Dividing both sides
of eqn (15) by $e^{\lambda}(1+Qe^{\lambda-\beta\bar{\epsilon}})^{\gamma-1}$ we
obtain
\begin{eqnarray}
1=e^{-\beta\bar{\epsilon}}\left(\frac{1+2e^{\lambda-\beta\bar{\epsilon}}}
{1+Qe^{\lambda-\beta\bar{\epsilon}}}\right)^{\gamma-1}
\end{eqnarray}
Rewrite eq.(8) as $2N/n=z_{gr}/B$ where $n=n_p+n_n$ to obtain
\begin{eqnarray}
\frac{2N}{n}=2+e^{-\lambda}\left(\frac{1+2e^{\lambda-\beta\bar{\epsilon}}}
{1+Qe^{\lambda-\beta\bar{\epsilon}}}\right)^{\gamma}
\end{eqnarray}
Using eq(15) the above relation leads to
\begin{eqnarray}
e^{\lambda}=\frac{n}{2(N-n)}e^{\beta\bar{\epsilon}\gamma/(\gamma-1)}
\end{eqnarray}
Define $x=e^{-\beta\bar{\epsilon}/(\gamma-1)}$.  Going back to eq(17) one
can now derive a simple solution for $x$:
\begin{eqnarray}
x=\frac{1}{2}\left[\frac{N-2n}{N-n}+\sqrt{\left(\frac{N-2n}{N-n}\right)^2+
2Q\frac{n}{N-n}}\right]
\end{eqnarray} 
Values of $\bar{\epsilon}$ and $e^{\lambda}$ from the definition of $x$ and
eq.(18).

Let us now go back to the partition function for the lattice as given by
eq.(3). where it is written as a product of the partition functions of the
$N/(\gamma+1)$ blocks.  If we simply $z_{gr}$ for each little block as
calculated above we will count twice the interaction between neighbouring
sites in different blocks.  For example, the binding energy between 1 and 6
(Fig. A.1) is included in $z_{gr}$(block 1) and it is included again in
$z_{gr}$(block 2).  We note that on the average there are $n/N$ particles at
each site and each block has $\gamma$ peripheral sites.  Thus when we evaluate
the partition function for the lattice, the partition function for each block
should be corrected by the multiplicative factor
\begin{eqnarray}
correction=e^{\frac{1}{2}\beta\bar{\epsilon}\gamma n/N}
\end{eqnarray}
We can now use $PV=kT\ln Z{gr}, V=N/\rho_0$, and $\ln Z_{gr}=N/(\gamma+1)\ln z_{gr}
$ to obtain
\begin{eqnarray}
P=\rho_0kT\frac{1}{\gamma+1}\ln z_{gr}
\end{eqnarray}
where $z_{gr}$ includes the correction factor.

In Fig. A2 we have drawn $P-V$ diagrams for nuclear matter for two
cases: (1) $\epsilon_{pn}=\epsilon_{pp}=-5.33$ MeV and 
(2)$\epsilon_{pn}=-5.33$ MeV
and $\epsilon_{pp}=0$ MeV.  In Bethe-Peierls approximation the $T_c$ in the
former case appears to be between 6 and 8 MeV and in the second case between 4
and 6 MeV.
 
\pagebreak

\begin{figure} 
\caption{ 
A comparison of the calculated values of $\tau$, $C_v$ and the second
moment $S_2$ in the one type of bond model (top panel) and two types of
bonds model (bottom panel).  Here as elsewhere $T_c\equiv 1.1275|\epsilon_{pn}|
\approx 6$ MeV.  The compound system has A=103,Z=45.  Notice the maxima and the
minimum shift to lower temperature when $\epsilon_{pp}$ is set to zero.  Also
$C_v$ is more sharply peaked.
}

\bigskip
\caption{ 
The same as above except that a higher freeze-out density is used.  The
number of lattice sites is still $7^3$.  Here A=171, Z=70.
}

\bigskip
\caption{ 
The top part compares the $\tau$ values extracted from a lattice gas 
calculation (dotted curve) with those extracted from a molecular dynamics
calculation which had no Coulomb (dashed curve) and one which had Coulomb
included in the molecular dymamics calculation (solid curve).  Molecular
dynamics without Coulomb gives results very similar to those of the lattice
gas model.  Here the effect of the Coulomb is small.  The number of protons
was 45. The lower part compares the second moments.
}

\bigskip
\caption{ 
Effect of Coulomb on $\tau$ and the second moment for a much larger
system; A=197, Z=79.  The minimum in $\tau$ and the maximum in $S_2$ shift
from about 5 MeV to 2.4 MeV because of Coulomb effect.  At some larger Coulomb 
field the minimum in $\tau$ will finally disappear.
}
\end{figure}

FIG. A1.  A square lattice is divided into blocks to illustrate the
Bethe-Peierls approximation.  See text for details.

FIG. A2. p-V diagram in the Bethe-peierls approximation with same bond strength
-5.33 MeV (top panel) between all particles and for the case where we
distinguish between like particle interaction and unlike particle interaction
(bottom panel). 

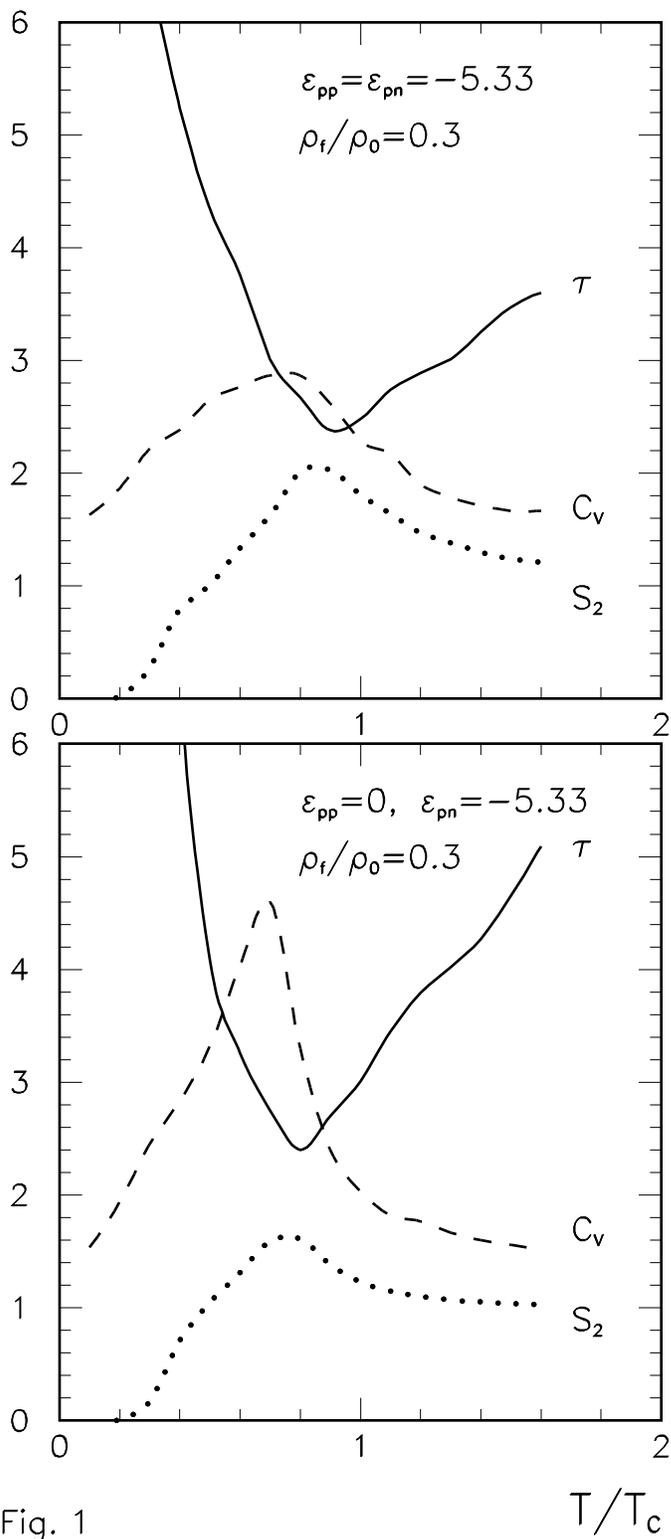
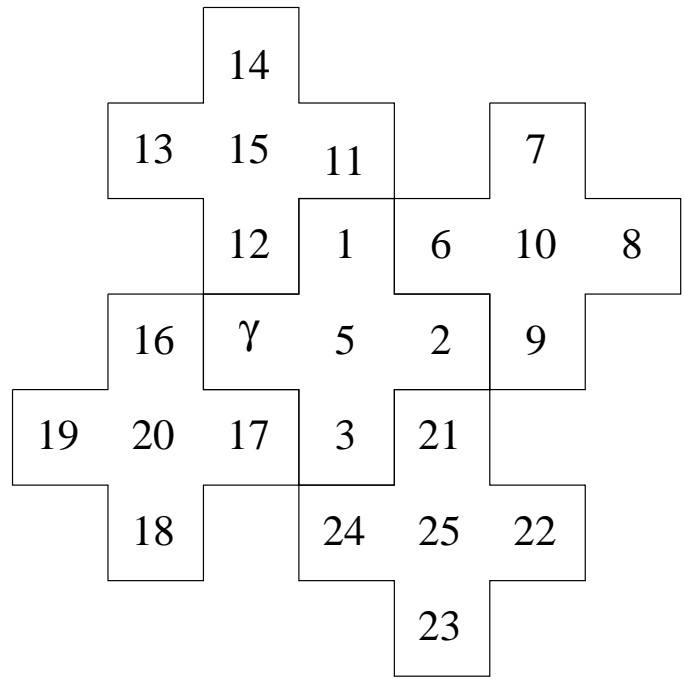
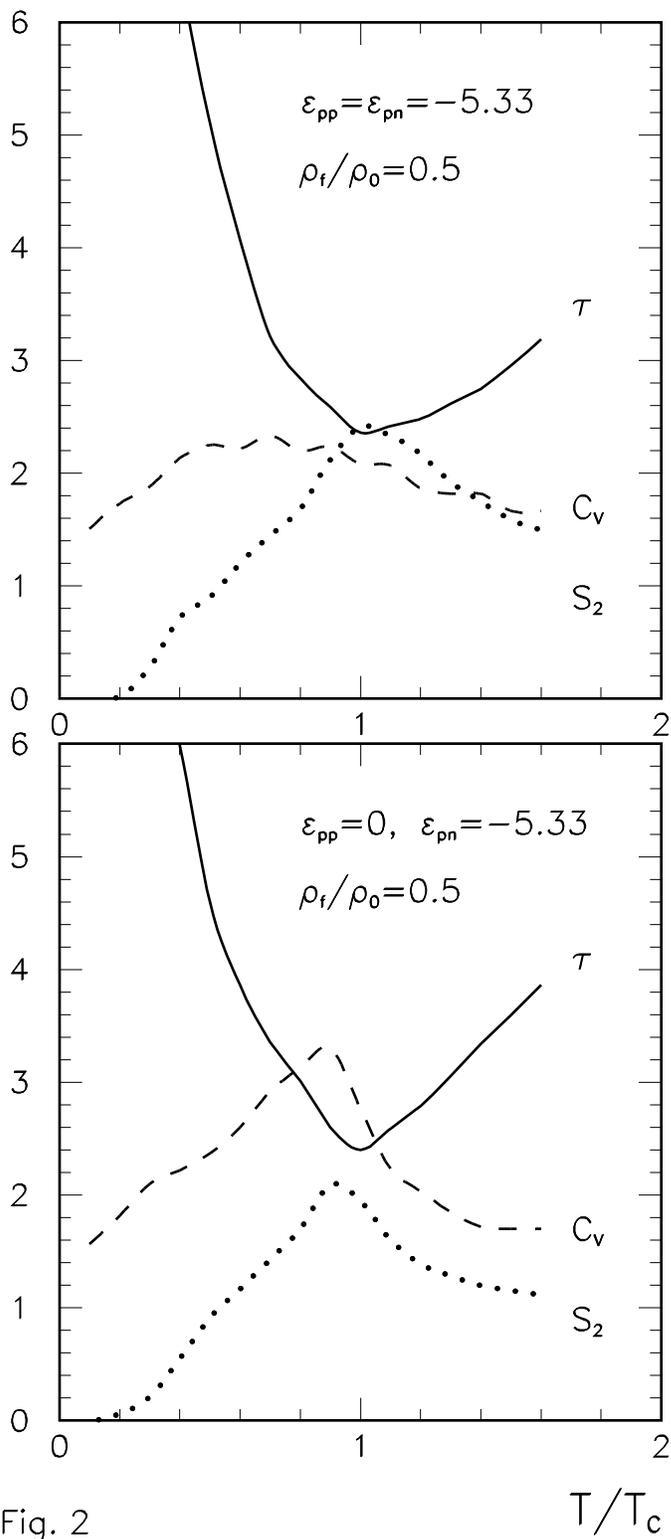
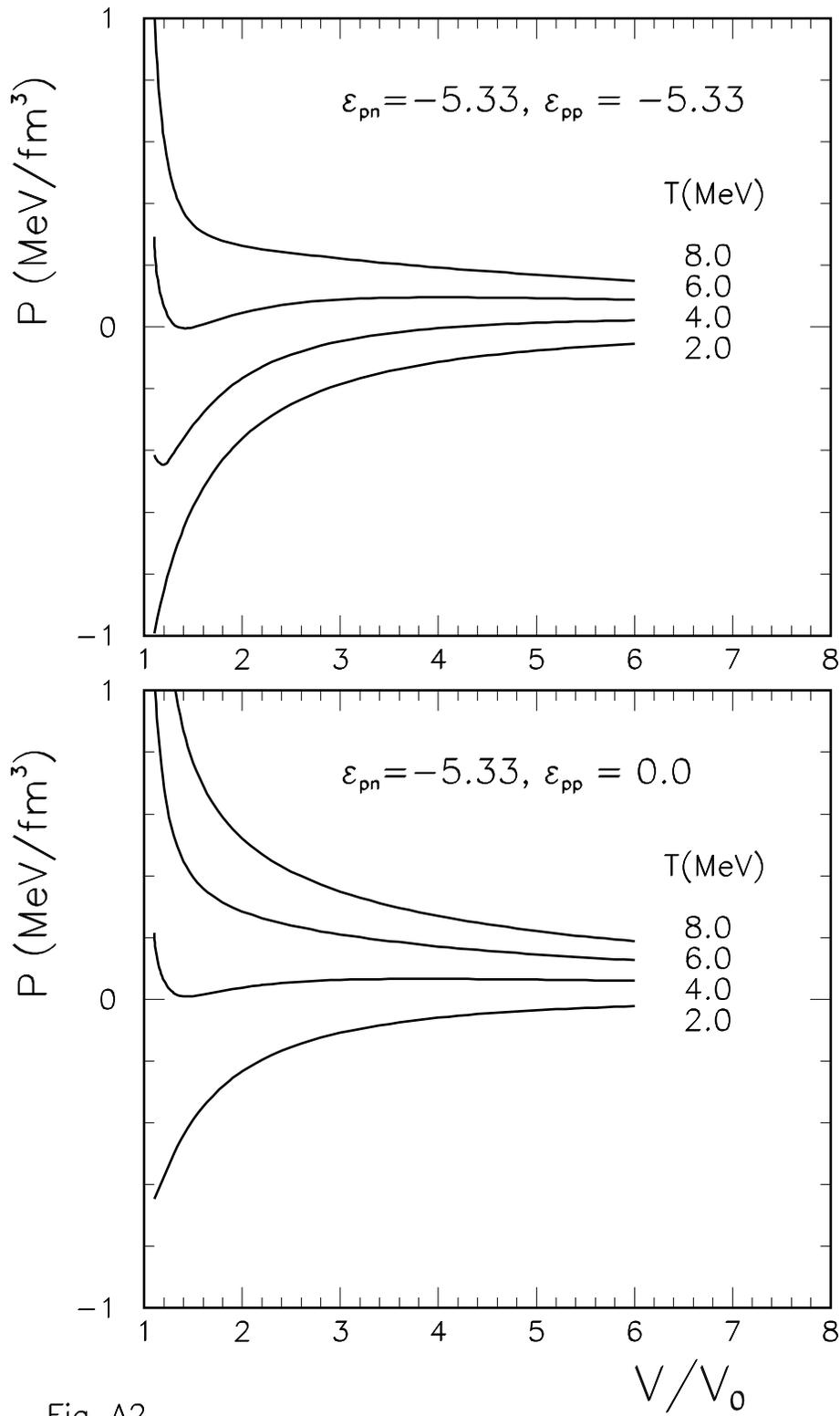
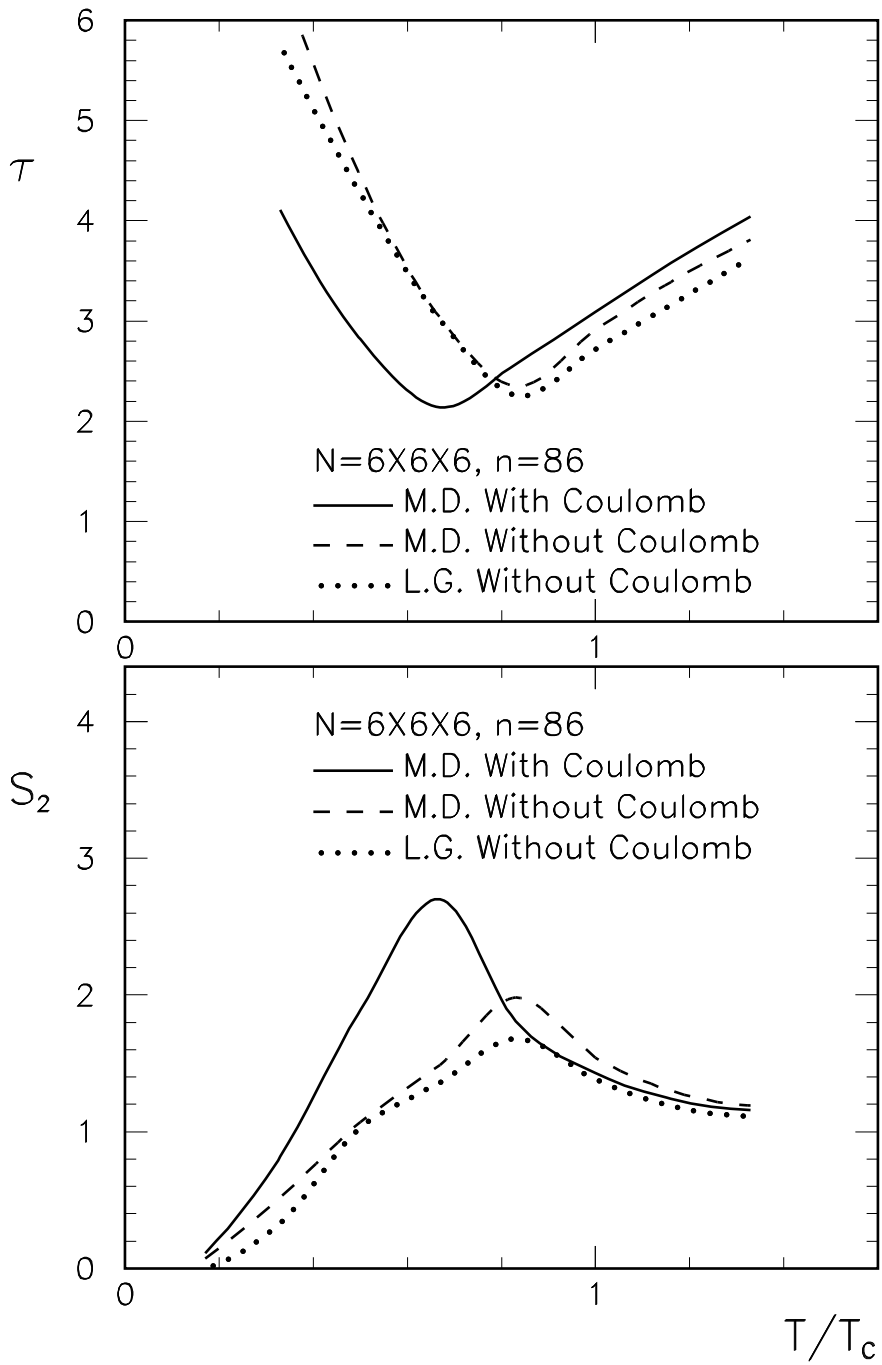
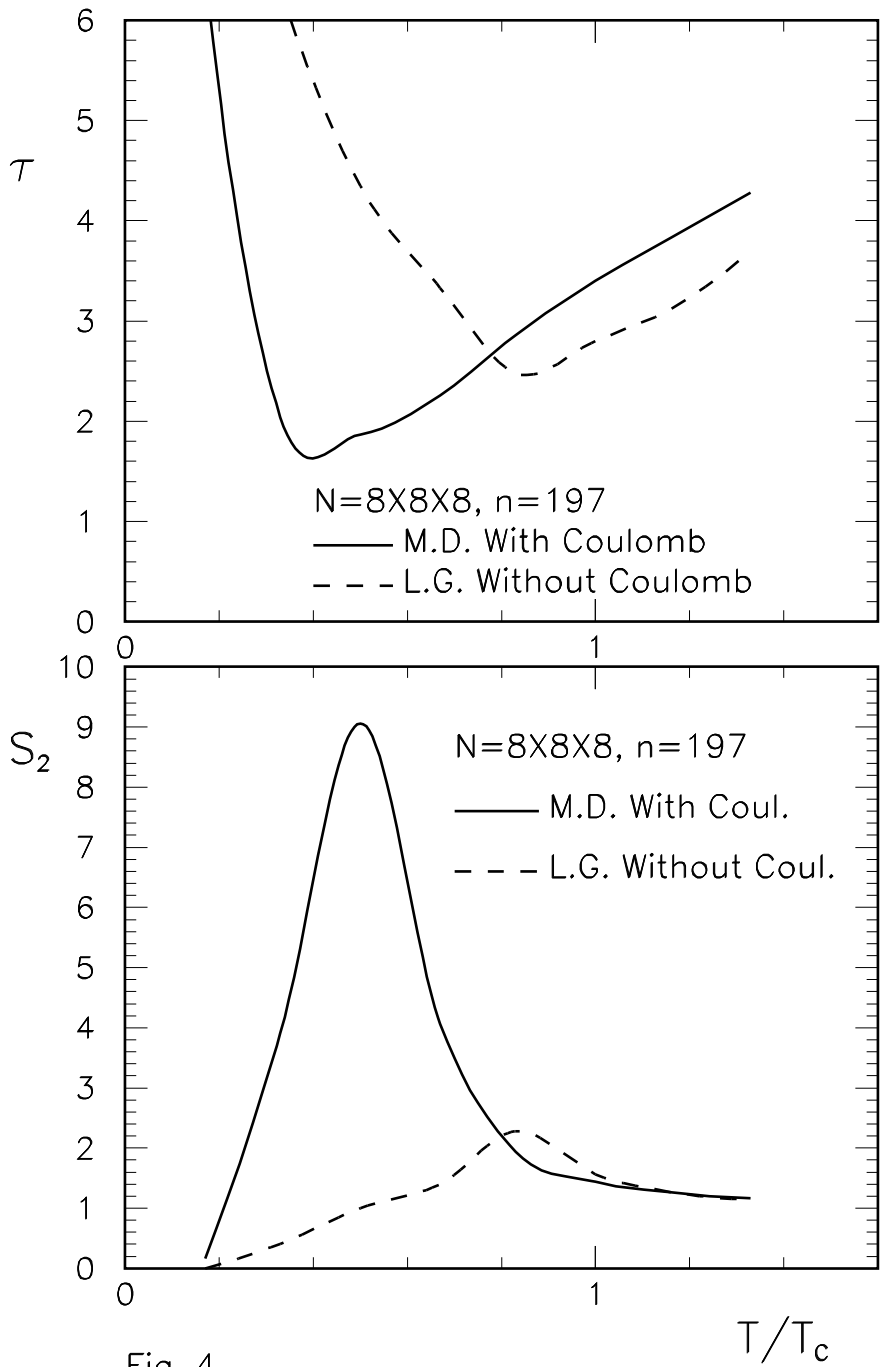
\end{document}